\documentstyle[sprocl,psfig]{article}

\def\be{\begin{equation}}
\def\ee{\end{equation}}
\def\bea{\begin{eqnarray}}
\def\eea{\end{eqnarray}}

\begin{document}
\begin{flushright}
  IC-HEP/00-02 \\
\end{flushright}

\title{
       FUTURE HIGH $\bf{Q^2}$ DEEP INELASTIC SCATTERING AT HERA
       }

\author{K. Long}

\address{
         Dept. of Physics, Blackett Laboratory, Imperial College London,
         Prince Consort Road, London SW7 2BW, United Kingdom            \\
         E-mail: K.Long@IC.AC.UK                                        
         \footnote{\sl
           Presented at the 8$^{th}$ International Workshop on Deep Inelastic
           Scattering, Liverpool, 25$^{th}$ - 30$^{th}$ April 2000 
                   }
         }


\maketitle\abstracts{ 
The luminosity of the electron-proton collider, HERA, will be
increased by a factor of five during the long shutdown starting September
2000. 
At the same time longitudinal lepton beam polarisation
will be provided for the collider experiments H1 and ZEUS. These far
reaching upgrades to the machine will be matched by upgrades to the
detectors. The result will be a unique facility for the study of the
structure of the proton and the nature of the strong and electroweak
interactions. The physics potential of the upgraded accelerator is
discussed here.
                       }

\section{Introduction}
The electron-proton collider HERA started operation in the summer of
1992. The proton beam energy was 820 GeV while the electron beam
energy was 26.7~GeV and was later raised to 27.5~GeV. 
In the years 1992-1997 H1 and ZEUS each collected a luminosity of 
$\sim$1 pb$^{-1}$ using electron beams and $\sim$50 pb$^{-1}$ 
using positron beams. 
These data have extended the kinematic 
range covered by deep inelastic scattering, DIS, measurements by two orders
of magnitude in both $Q^{2}$, the four-momentum transfer squared, and $x$,
the fraction of the proton four-momentum carried by the struck
quark. 
These data have been used to dertermine the proton structure function,
make measurements which 
test the electroweak Standard Model, SM, and the theory of the strong 
interactions, QCD, in both neutral current, NC, and charged current, CC, DIS. 
Jet analyses in DIS and photoproduction have been used to address fundamental
issues in QCD. The observation of diffraction in DIS has led to a
careful investigation of the transition from the kinematic region in
which perturbative QCD is valid to the region where phenomenological
models based on Regge theory must be applied (see for example \cite{Cooper} 
and \cite{Abramovicz} and references therein). 

During the running period August 1998 to April 1999 $\sim$20 pb$^{-1}$ 
of $e^{-}p$ data were delivered with a proton beam energy of 920 GeV. 
HERA is now delivering $e^+p$ collisions with a proton beam energy 
of 920~GeV.
By the end of running in September 2000 H1 and ZEUS will each have an 
$e^+p$ data set of $\sim 100$~pb$^{-1}$.
The data
collected by H1 and ZEUS will be used to study the dependence of the
NC and CC DIS cross sections on the charge of the lepton beam.

The HERA experiments will continue to take data until September 2000 when a
long, 9 month, shutdown is scheduled. The shutdown will be used to
upgrade the HERA accelerator and the collider detectors. 
The HERA luminosity will be increased by a factor of five and
longitudinal lepton beam polarisation ($\sim 70\%$) will be provided for ZEUS
and H1.
Over a six year running period it is anticipated that a total
luminosity of 1000 pb$^{-1}$ will be delivered \cite{Schneekloth}.
The physics motivation for this major upgrade programme is discussed
in detail in reference \cite{Ingelman}.

\section{Physics at HERA after the Upgrade}

Following the HERA upgrade the proton will be probed using each of the
four possible combinations of lepton beam charge and polarisation. The
combination of high luminosity and polarisation will lead to a rich
and diverse programme of measurements which can only be sketched below
using a few examples. 

\subsection{Proton Structure}

The large data volume will allow $F_{2}^{\rm NC}$ to be extracted with an 
accuracy of $\sim$3\% over the kinematic range $2 \times 10^{-5}<x<0.7$ 
and $2 \times 10^{-5} <Q^{2} <5 \times 10^{4}\,{\rm GeV}^{2}$ \cite{Botje}. 
If QCD evolution 
codes which go beyond next to leading order become available and a careful
study of the dependence of the systematic errors on the kinematic
variables is made it will be possible to determine $\alpha_{\rm S}$ from the 
scaling violations of $F_{2}^{\rm NC}$ with a precision of $\leq 0.003$. The 
gluon distribution will also be determined from such a fit with a
precision of $\sim$ 3\% for $x = 10^{-4}$ and $Q^{2} = 20\,{\rm GeV}^{2}$. 

The combination of high luminosity and high charm tagging efficiency 
transforms the measurement of the charm contribution to 
$F_{2}^{\rm NC}, F_{2}^{cc}$ \cite{Daum}. 
The 
precision will be sufficient to allow a detailed study of the charm 
production cross section to be made. The lifetime tag provided by the silicon 
micro-vertex detector allows the tagging of $b$-quarks and the
determination of the ratio of the beauty contribution to $F_{2}^{\rm NC}$,
$F_{2}^{bb}$, to $F_{2}^{cc}$.

In the quark parton model CC DIS is sensitive to specific quark
flavours. The $e^{+}p$ CC DIS cross section is sensitive to the $d$- and 
$s$-quark parton densities and the $\bar{u}$- and $\bar{c}$-anti-quark 
densities, while the $e^{-}p$ CC DIS cross section is sensitive to the 
$u, c, \bar{d}$ and $\bar{s}$ parton density functions. With the large CC 
data sets expected following the upgrade it will be possible to use 
$e^{\pm}p$ CC data to determine the $u$- and $d$- quark densities. Further, by 
identifying charm in CC DIS it will be possible to determine the strange 
quark contribution to the proton structure function $F_{2}^{\rm NC}$ with an 
accuracy of between 15\% and 30\% \cite{Lamberti}.

\subsection{Tests of the Electroweak Standard Model}

The high luminosity provided by the upgrade will allow access to low cross
section phenomena such as the production of real $W$-bosons. The SM
cross section for the process $ep \rightarrow eWX$ is $\sim$~1~pb 
\cite{WCross} which, combined with an acceptance of $\sim30\%$, gives a 
sizeable data sample for a luminosity of 1000 pb$^{-1}$. The production of 
the $W$-boson at HERA is sensitive to the non-abelian coupling $WW\gamma$ 
\cite{Noyes}. The sensitivity of HERA to non-SM couplings is comparable to 
the sensitivities obtained at LEP and at the Tevatron and complementary in 
that at HERA is predominantly sensitive to the $WW\gamma$ vertex,
independent of assumptions about the nature of the $WWZ$ vertex.

The full potential of electroweak tests at HERA will be realised
through measurements using polarised lepton beams \cite{Cashmore}.
Within the SM NC and CC DIS cross sections may be written in terms of
$\alpha$, $M_{W}$ and $m_{t}$ together with the mass of the $Z$ boson, 
$M_{Z}$, and the mass of the Higgs boson, $M_{H}$. In order to test the 
consistency of the theory we may fix the values of $\alpha$ and $M_{Z}$ to 
those obtained at LEP or elsewhere and use measurements of the NC and CC DIS 
cross sections to place constraints in the $M_{W}$, $m_{t}$ plane for fixed 
values of $M_{H}$. The SM is consistent if the values of the parameters 
$M_{W}$ and $m_{t}$ obtained agree with the values determined in other 
experiments. 
Combining NC and CC data 
corresponding to a luminosity of 1000 pb$^{-1}$, recorded with a
lepton beam polarisation of 70\%, with a top mass measurement 
from the Tevatron with a precision of $\sim\pm5$ GeV yields a
measurement of $M_{W}$ with an error of $\sim60$ MeV \cite{Beyer}. 
\begin{figure}
  \centerline{\psfig{figure=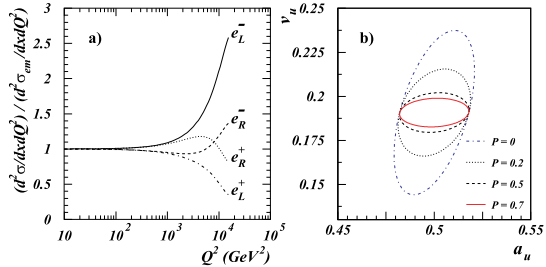,width=.6\textwidth}}
  \caption{
    \label{Fig:Fig9}
    (a) Ratio of the NC DIS cross section to the cross section obtained
    when only single photon exchange is included as a function of $Q^{2}$ 
    at $x=0.2$. (b) Sensitivity of the errors on the $u$-quark
    coupling to the $Z$ to the lepton beam polarisation $P$. One
    standard deviation contours are shown for fits in which the
    $u$-quark couplings are allowed to vary while the $d$-quark
    couplings are held fixed at their SM values~$^{12}$.
  }
\end{figure}

The sensitivity of NC DIS to lepton beam polarisation is shown in
figure \ref{Fig:Fig9}(a). The figure shows the ratio of the full NC
cross section to the cross section obtained in the single photon
exchange approximation.
The strong 
polarisation dependence of the NC cross section can be used to extract the NC 
couplings of the light quarks. In such an analysis the CC cross section may 
be used to reduce the sensitivity of the results to uncertainties in the
PDFs \cite{Klanner}. The precision of the results obtained depends
strongly on the degree of polarisation of the lepton beam as shown in
figure \ref{Fig:Fig9}(b). The
figure shows the anticipated error on the vector and axial-vector
couplings of the $u$-quark, $v_{u}$ and $a_{u}$ respectively, obtained in a 
fit in which $v_{u}$ and $a_{u}$ are allowed to vary while all other 
couplings are fixed at their SM values. With a luminosity of 250 pb$^{-1}$ 
per charge, polarisation combination and taking the vector and axial-vector
couplings of the $u$- and $d$-quarks as free parameters gives a precision
of 13\%, 6\%, 17\% and 17\% for $v_{u}$, $a_{u}$, $v_{d}$ and $a_{d}$ 
respectively. By comparing these results with the NC couplings of the $c$- and 
$b$-quarks obtained at LEP a stringent test of the universality of the NC 
couplings of the quarks will be made. 


\end{document}